\documentclass[9pt,twocolumn,twoside]{opticajnl}
\journal{opticajournal} % use for journal or Optica Open submissions

\setboolean{shortarticle}{false}

\usepackage{tikz}
\usepackage{bm}
\usepackage{multirow}
\usepackage{pifont}
\usepackage{mathtools}
\usepackage{pgfplots}
\usepackage{wasysym}
\usepackage{cleveref}
\usepackage{siunitx}
\usepackage{tabu}
\usepackage{adjustbox}

\usepgfplotslibrary{groupplots,dateplot}
\pgfplotsset{compat=newest}
\usetikzlibrary{patterns,shapes,arrows,arrows.meta,chains,fit,positioning,calc,3d}

\def\centerarc[#1](#2)(#3:#4:#5)% [draw options] (center) (initial angle:final angle:radius)
{
    \draw[#1] ($(#2)+({#5*cos(#3)},{#5*sin(#3)})$) arc (#3:#4:#5);
}
\def\connectionline[#1](#2)(#3)(#4:#5)% [draw options] (start) (end) (first angle:second angle:radius)
{
    \draw[#1] ($(#2)+({#5*cos(#4)},{#5*sin(#4)})$) -- ($(#3)+({#5*cos(#4)},{#5*sin(#4)})$);
}

\newcommand{\cmark}{\ding{51}}%
\newcommand{\xmark}{\ding{55}}%

\graphicspath{{images/}}
\colorlet{lcfree}{black}
\colorlet{lcnorm}{black}
\colorlet{lccong}{black}

\newcommand{\fig}{Fig.~}
\newcommand{\tab}{Table~}
\newcommand{\T}{^\text{T}}

\newcommand{\calcoordM}{m}
\newcommand{\calcoordN}{n}
\newcommand{\calcoord}{\calcoordM_\channel, \calcoordN_\channel}
\newcommand{\descoordX}{x}
\newcommand{\descoordY}{y}
\newcommand{\descoord}{\descoordX, \descoordY}

\newcommand{\numchannels}{N}
\newcommand{\imagewidth}{S_w}
\newcommand{\imageheight}{S_h}

\newcommand{\channel}{c}
\newcommand{\centercam}{m}
\newcommand{\ssimconst}{g}

\newcommand{\homography}{\bm{H}}

\newcommand{\calibrated}{C}
\newcommand{\disparity}{D}
\newcommand{\hsimage}{H}
\newcommand{\groundtruth}{G}
\newcommand{\hsmasked}{K}
\newcommand{\mask}{M}

\newcommand{\hsimagewarp}{W}

\newcommand{\spectrum}{\bm{s}}

\newcommand{\mean}{\mu}
\newcommand{\variance}{\sigma}

\fancypagestyle{firststyle}
{
	\fancyhf{}
	\fancyfoot[L]{\copyright 2024 Optica Publishing Group. One print or electronic copy may be made for personal use only. Systematic reproduction and distribution, duplication of any material in this paper for a fee or for commercial purposes, or modifications of the content of this paper are prohibited.}
}

\title{High-Resolution Hyperspectral Video Imaging Using a Hexagonal Camera Array}%

\author[1, *]{Frank Sippel}%
\author[1]{Jürgen Seiler}%
\author[1]{André Kaup}%

\affil{Chair of Multimedia Communications and Signal Processing, Friedrich-Alexander-Universität Erlangen-Nürnberg (FAU), Cauerstr. 7, 91058 Erlangen, Germany}
\affil[*]{frank.sippel@fau.de}

\begin{abstract}
Retrieving the reflectance spectrum from objects is an essential task for many classification and detection problems, since many materials and processes have a unique spectral behaviour.
In many cases, it is highly desirable to capture hyperspectral images due to the high spectral flexibility.
Often, it is even necessary to capture hyperspectral videos or at least to be able to record a hyperspectral image at once, also called snapshot hyperspectral imaging, to avoid spectral smearing.
For this task, a high-resolution snapshot hyperspectral camera array using a hexagonal shape is introduced.
The hexagonal array for hyperspectral imaging uses off-the-shelf hardware, which enables high flexibility regarding employed cameras, lenses and filters.
Hence, the spectral range can be easily varied by mounting a different set of filters.
Moreover, the concept of using off-the-shelf hardware enables low prices in comparison to other approaches with highly specialized hardware.
Since classical industrial cameras are used in this hyperspectral camera array, the spatial and temporal resolution is very high, while recording 37 hyperspectral channels in the range from 400 nm to 760 nm in 10 nm steps.
As the cameras are at different spatial position, a registration process is required for near-field imaging, which maps the peripheral camera views to the center view.
It is shown that this combination using a hyperspectral camera array and the corresponding image registration pipeline is superior in comparison to other popular snapshot approaches.
For this evaluation, a synthetic hyperspectral database is rendered.
On the synthetic data, the novel approach outperforms its best competitor by more than 3 dB in reconstruction quality.
This synthetic data is also used to show the superiority of the hexagonal shape in comparison to an orthogonal-spaced one.
Moreover, a real-world high resolution hyperspectral video database with ten scenes is provided for further research in other applications.
The code and data are available at \url{https://github.com/FAU-LMS/HAHSI}.
\end{abstract}

\setboolean{displaycopyright}{false} % Do not include copyright or licensing information in submission.

\begin{document}
\taburulecolor{black}

\maketitle

\thispagestyle{firststyle}

\section{Introduction}
\label{sec:introduction}
Classical cameras, like a DSLR or a smartphone, record scenes in red, green and blue, because they try to capture the required information for the human visual system.
For other purposes, it is useful to record even more channels and possibly also in other wavelength areas, e.g., in the infrared (IR).
This concept is called multispectral imaging, where typically six to twelve channels are captured.
Hyperspectral imaging on the other hand aims at sampling the light spectrum within certain spectral bounds.
Hence, usually the same or at least a very similar bandwidth is used for the different bands.
Moreover, the sampling distance is kept as uniformly as possible.
Thus, hyperspectral imaging is rather useful if the spectrum is not yet known or there is a need to capture slight variations in the spectrum like small spectrum shifts.

Hyperspectral imaging has diverse applications.
It can be used to separate different types of plastic~\cite{garaba_plastic_2018}, in agriculture~\cite{williams_maize_2016}, to detect traces in forensics~\cite{edelman_forensics_2012}, for object tracking~\cite{xiong_tracking_2020}, and in medicine to detect tumors~\cite{han_medicine_2016}.
Depending on the scenario, a high spatial and temporal resolution may be required in all of these applications.
However, there is a lack of high resolution hyperspectral snapshot cameras so far.

\begin{table*}[t]
    \vspace*{-0.5cm}
	\centering
	\renewcommand{\arraystretch}{1.5}
    \caption{Overview over different multi- and hyperspectral cameras and their properties.}
    \label{tab:cams}
    \begin{tabular}{cccccc}
        Camera                                          & Spatial resolution    & Filter flexibility    & Snapshot  & Reconstruction required   & Depth     \\ \hline
        Pushbroom~\cite{gomez-chova_correction_2008}    & \cmark                & \xmark                & \xmark    & \xmark                    & \xmark    \\
        Filter wheel~\cite{koenig_wheel_1998}           & \cmark                & \cmark                & \xmark    & \xmark                    & \xmark    \\
        Fiber Bundles~\cite{gat_4dis_2006}              & \xmark                & \xmark                & \cmark    & \cmark                    & \xmark    \\
        CTI~\cite{descour_cti_1997}                     & \xmark                & \xmark                & \cmark    & \cmark                    & \xmark    \\
        Beamsplitter~\cite{matchett_beamsplitter_2007}  & \cmark                & \xmark                & \cmark    & \xmark                    & \xmark    \\
        Lens Array~\cite{shogenji_lensarray_2004}       & \xmark                & \xmark                & \cmark    & \xmark                    & \xmark    \\
        MSFA~\cite{monno_msfa_2015}                     & \cmark                & \xmark                & \cmark    & \cmark                    & \xmark    \\
        CASSI~\cite{gehm_cassi_2007}                    & \cmark                & (\cmark)              & (\cmark)  & \cmark                    & \xmark    \\
        CAMSI~\cite{genser_camsi_2020}                  & \cmark                & \cmark                & \cmark    & \cmark                    & \cmark    \\
        Proposed HAHSI                                  & \cmark                & \cmark                & \cmark    & \cmark                    & \cmark    \\
	\end{tabular}
\end{table*}

When imaging a scene hyperspectrally, a three-dimensional data cube with two spatial dimensions and one spectral dimension has to be captured.
A hyperspectral camera needs to unfold this three-dimensional data cube using 2D grayscale sensors.
Hyperspectral cameras can be divided in scanning and snapshot devices.
For scanning methods, one dimension of this hyperspectral data cube is often unfolded over time.
A filter wheel\cite{koenig_wheel_1998} unrolls the spectral dimension over time by rotating a wheel with filters in front of a single camera.
Similarly, Pushbroom sensors\cite{gomez-chova_correction_2008} scan one of the two spatial dimensions over time, e.g., by placing the objects to scan on a conveyor belt.
These methods are fine if there are no restrictions on the imaging time, for example, because of a static scene.
However, as soon as moving objects or spectral temporal changes are involved, these scanning techniques will produce blurred images and are hence suboptimal.

Snapshot hyperspectral cameras have the goal of imaging the scene at once.
Thus, these snapshot hyperspectral devices are implicitly also able to capture videos.
Different approaches and their advantages are summarized in \tab\ref{tab:cams} and are explained in detail in the following.
One of the earlier approaches for this is by employing fiber bundles~\cite{gat_4dis_2006}, which can be arranged in a 2D grid.
Afterwards, these fiber bundles are rearranged to a one dimensional row and the light is sent through a disperser.
Finally, the 2D grayscale sensor senses the flattened spatial dimensions in one direction and the spectrum in the other direction.
Hence, the spatial resolution of this approach is very limited.
Another earlier approach is by using Computed Tomography Imaging (CTI)~\cite{descour_cti_1997}, where a 2D dispersion pattern is used.
Tomographic reconstruction is then able to estimate the data cube from multiple projections and view angles.
Since these projections fall onto the same sensor, again the spatial resolution of these approach is rather limited.
The beamsplitter approach~\cite{matchett_beamsplitter_2007} uses semitransparent mirrors, which reflect a specific spectral area, while the other spectral parts are passing.
Then, multiple sensors can pick up these different images.
There, the problem is that the semitransparent mirrors are not perfect and therefore multiple of these beamsplitters in a row lead to a very weak signal in the end.
Furthermore, lens arrays~\cite{shogenji_lensarray_2004} are possible, where an array of lenses project the image onto different parts of the sensor.
These different parts of the sensor are covered by individual bandpass filters.
Again, only a single sensor is used, which limits the spatial resolution of the resulting hyperspectral image.
Approaches based on compressed sensing~\cite{gehm_cassi_2007} are also popular, where a coded mask occludes part of the image.
Afterwards, a disperser is applied to this masked data cube, where the sum from different pixels and different bands is then sensed by a 2D grayscale sensor.
This concept is called Coded Aperture Snapshot Spectral Imager (CASSI).
Unfortunately, usually a set of frames is necessary to retrieve a proper hyperspectral data cube, which limits its snapshot capability.
This concept can be used in conjunction with tunable filters leading to filter flexibility to some extent~\cite{cassi_flex_2018}.
A very popular approach is extending the classical Bayer pattern~\cite{bayer_1976} for RGB imaging to MultiSpectral Filter Arrays (MSFA)~\cite{monno_msfa_2015}.
There, the filter pattern is periodically repeated on the grayscale sensor.
This approach needs a demosaicing process similar to the demosaicing process of RGB Bayer patterns.
Once again, only a single sensor in conjunction with the demosaicing process is used which limits the underlying spatial resolution.
Of course, this concept can also be used for hyperspectral imaging by using the corresponding filters within the filter array.
Finally, multi camera approaches use multiple spatially distributed cameras with different filters mounted in front of them.
This approach has been successfully applied to near-field multispectral imaging~\cite{genser_camsi_2020}.
In this paper, we progress this camera array approach to hyperspectral imaging by using 37 cameras arranged in a hexagonal grid along the necessary image signal processing.
We are using off-the-shelf components for this hexagonal array for hyperspectral imaging, which enables a high spatial and temporal resolution and a lower price point than the presented methods which all need highly specialized hardware.
Moreover, depending on the requirements, the cameras, the lenses and the filters can be exchanged, and therewith enabling different spatial and temporal resolution as well as a different spectral range if the filters are exchanged.

Camera arrays for hyperspectral imaging~\cite{cam_array_notch_2022} also have been used for art paintings~\cite{cam_array_art_2022} and agriculture~\cite{cam_array_wheat_2021}, however, they do not aim at general purpose hyperspectral imaging.
They only record the scene in one depth plane, thus only a calibration procedure is necessary.
In contrast, for near-field hyperspectral imaging, the challenge is that different objects are placed at different depth layers.
Therefore, the first step after calibrating the camera array is to estimate a depth map in a cross spectral way.
This disparity map then can be used to warp the peripheral cameras to the center view.
Unfortunately, due to depth occlusions the center camera sees pixels which are not visible to all of the peripheral cameras.
Consequently, these areas need to be detected for each peripheral view.
In a final step, these areas of the different peripheral views are reconstructed by exploiting the structure of the center view, which captures the scene in another spectral band.
Hence, an advanced signal processing pipeline is required to register hyperspectral images from such a camera array.

In this paper, a hyperspectral camera array with 37 cameras is presented along with its required image processing pipeline.
Moreover, a corresponding synthetic hyperspectral database is created, where a digital twin of this camera array records scenes virtually.
This database is used to show the superiority of the presented setup in comparison to filter array and coded aperture approaches.
Finally, a real-world high-resolution hyperspectral video database with ten scenes is introduced, which can be used for diverse research applications.

\section{Hyperspectral Camera Array}
\label{sec:camera_array}
\begin{figure}[t]
    \centering
    \input{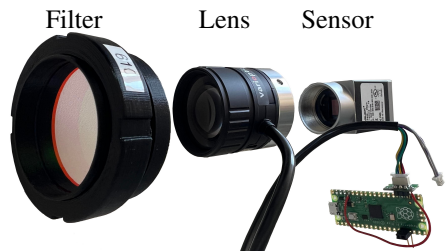}
    \caption{A single sensor-lens-filter combination.}
    \label{fig:cam}
\end{figure}

In the following, the \underline{H}exagonal \underline{A}rray for \underline{H}yper\underline{S}pectral \underline{I}maging (HAHSI) is introduced.
It consists of 37 sensor-lens-filter combinations arranged in a hexagonal grid.
One sensor-lens-filter combination is depicted in \fig\ref{fig:cam}.
The following subsections describe all of the single elements as well as the arrangement and the communication structure within this array.

\subsection{Camera sensors}

The utilized cameras are Basler ace acA2440-20gm industrial cameras.
They have a resolution of 2448 \texttimes\ 2048 pixels with a sensor size of 8.4 mm \texttimes\ 7.1 mm.
Hence, the squared pixels have a size of 3.45 \texttimes\ 3.45 \textmu m.
The underlying sensors are Sony IMX264 using a global shutter.
At full resolution they can record a scene at up to 23 frames per second (FPS).
However, for a smaller resolution, the frame rate can be increased.
For example, for a resolution of 600 \texttimes\ 400 pixels, they achieve a frame rate of 173 FPS.
The industrial cameras transmit their image data using Ethernet with a maximum data rate of 1 Gbps, which is the main reason for the frame rate limit.
Hence, if only a small part of the images is necessary, a very high temporal resolution is possible under the assumption of a well illuminated scene.
A convenient feature of these cameras is the support of Power over Ethernet (PoE).
Thus, they can be powered via ethernet using a PoE-capable switch and do not need extra power wiring.

\subsection{Lenses}

The next essential component of such a setup are lenses.
Since 37 cameras are employed, it is cumbersome to set focus and aperture manually like in~\cite{genser_camsi_2020}.
Furthermore, motorized lenses to be able to control focus via software would dramatically increase power consumption of the array and increase the baseline between two cameras due to the additional hardware.
Thus, liquid lenses are used, specifically Corning C-C-39N0-160, which have a fixed aperture of f/2.8 and have an I2C interface to control focus.
The basic principle of liquid lenses is electrowetting~\cite{electrowetting}.
There, the shape of a drop of some liquid can be controlled by applying a voltage to the liquid and a substrate.
Between the liquid and the substrate, an isolator is mounted such that they cannot get in contact.
As soon as a voltage is applied, the negative charge of the liquid is attracted to the positive charge of the substrate.
Due to this force, the shape of the drop changes.

This principle can be exploited for liquid lenses, where two immiscible liquids are used to achieve the same effect.
Again, the shape of the drop of one of the liquids changes, which then modifies the optical path through the lens.
Thus, the focus can be set to different depths.
The big advantage is that no mechanics are necessary to control the focus, since the voltage can be easily set by software.
Hence, the required power consumption also stays lower.
Furthermore, the focus can be changed much quicker than with a conventional lens.

\begin{figure}[t]
    \centering
    \input{tikz/arrangements.tikz}
    \caption{Hexagonal arrangement for the hyperspectral camera array, including corresponding wavelengths on the left. On the right, HAHSI is depicted.}
    \label{fig:arrangements}
\end{figure}

\subsection{Filters}

To cut out the desired spectral bands, filters are required.
In the case of HAHSI, the filters record the spectral area from 400 nm to 760 nm in 10 nm steps, thus resulting in $\numchannels=37$ bands. % Maybe plot
The bandwidth of these filters is 10 nm.
Artifex dielectric interference filters with optical density 4 (OD4) blocking and a diameter of 40 mm are used.
OD4 states that less than 0.01 percent of light passes through the filter in the blocked wavelength areas.
Since the lenses do not have a filter holder, a custom 3D printed filter holder is attached to the lenses.
With this filter holder, the full lens is covered by the corresponding bandpass filter.

\subsection{Arrangement}

These sensor-lens-filter combinations are arranged in a hexagonal grid as shown in \fig\ref{fig:arrangements}.
The big advantage of the hexagonal shape is that the array can be built in a more compact way.
Thus, the maximum baseline between the center camera and the outer cameras is reduced, which leads to a bigger overlap and a smaller disparity in the resulting images.
Moreover, using this concept the center camera has six direct neighbors, which is an advantage for the signal processing explained in Section~\ref{sec:pipeline}.
The superiority of the hexagonal shape will be numerically evaluated in Section~\ref{subsec:shape_eval}.

\begin{figure}[t]
    \centering
    \input{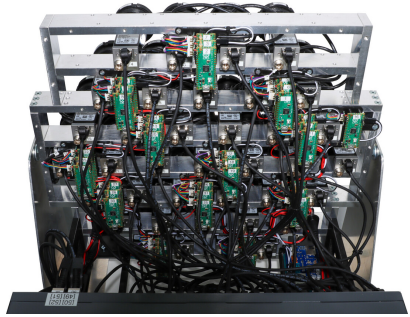}
    \caption{The communication backplane. Ethernet and I2C are used to communicate to the camera and lenses, respectively.}
    \label{fig:communication}
\end{figure}

\subsection{Communication}

\begin{figure}[t]
    \centering
    \input{tikz/communication_flow.tikz}
    \caption{The communication flow within HAHSI. The images are passed from the cameras to the PC. There, the images are analysed to determine focus values. These focus values are then transmitted to the liquid lenses.}
    \label{fig:communication_flow}
\end{figure}

The communication backplane is depicted in \fig\ref{fig:communication} and the corresponding flow diagram described in the following in \fig\ref{fig:communication_flow}.
The communication of the cameras with the controlling PC is done using a PoE-capable switch, which is able to power the cameras as well.
An FS S3400-48T4SP switch is used, which has 48 Gigabit ports to communicate with the cameras.
The data rate that can be maximally produced by the cameras is 37~Gbit/s.
To transport this amount of data to the PC, the switch has 4 SFP+ uplinks, which can transmit 40~Gbit/s in total.

The communication to the liquid lenses is less straightforward.
Unfortunately, every liquid lens has the same I2C address.
Thus, they cannot be easily connected to an I2C bus.
To cope with this behaviour, every liquid lens is connected to a Raspberry Pi Pico by I2C.
The Raspberry Pi Picos are programmed to listen to different I2C addresses on the bus and pass through the received values to the liquid lens.
Note that apart from a single lens, three picos are stacked on top of each other.
The bus itself is fed by a single Arduino Uno, which retrieves the focus values for all liquid lenses via ethernet.
Hence, the Arduino is also connected to the switch such that the controlling PC can set the focus of the individual cameras.
To enhance stability of the I2C communication, the bus is branched in two main veins, instead of one large bus that travels to every Pico.

\subsection{Synchronization}

\begin{figure}[t]
    \centering
    \input{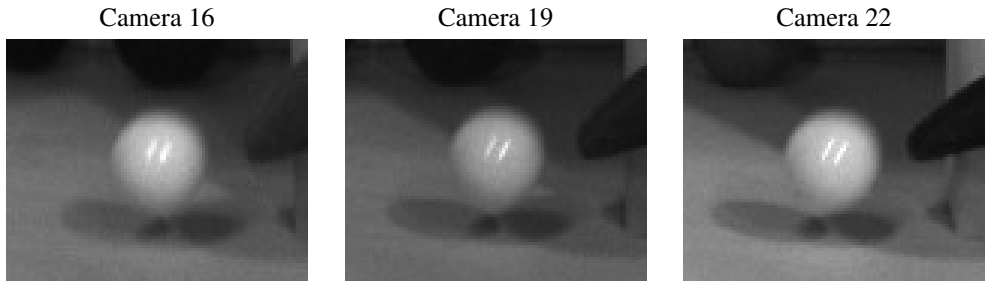}
    \caption{Cameras are synchronized via PTP. The images depict the same frame of a moving ball from different cameras. Camera 19 is the center camera of the camera array.}
    \label{fig:sync}
\end{figure}

A final important aspect of HAHSI is synchronization between cameras.
When the cameras are not synchronized, they record movement at different points in time, which results in objects being at different positions in different spectral bands.
Thus, the cameras should start to capture images at the exact same moments.
Fortunately, the cameras support the Precision Time Protocol (PTP)~\cite{ptp_2008}.
Here, one camera acts as grandmaster and yields its clock to all other cameras, which are slaves.
Since this time stamp needs to be transmitted over the network with differing transportation times, this procedure is a transient process.
In the end, all cameras have nearly the exact same time stamp with an accuracy of a few hundred nanoseconds.
Afterwards, the cameras do not need to communicate anymore, since time stamps when to capture images are predefined.
Hence, the synchronization needs to be executed once, when the array is started.
\fig\ref{fig:sync} shows three different bands and thus cameras synchronized by PTP.
As long as the exposure time stays the same, the resulting images are highly synchronized.

\section{Registration Pipeline}
\label{sec:pipeline}
\begin{figure*}[t]
    \centering
    \input{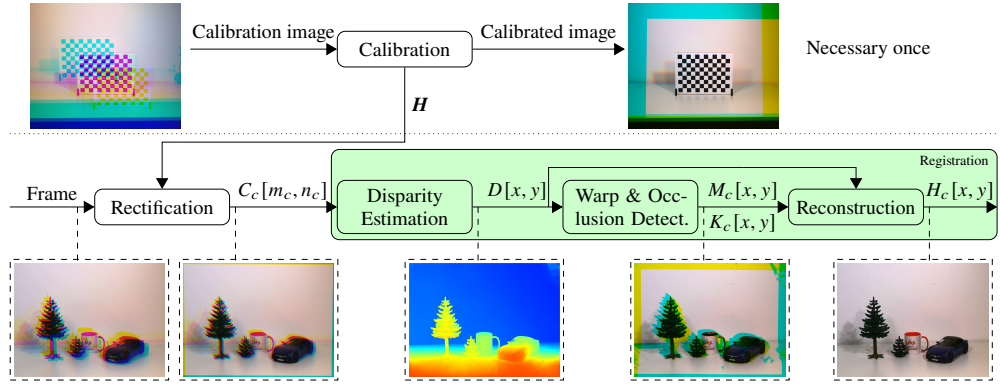}
    \caption{The registration image processing pipeline to warp the peripheral views to the center view assuming undistorted lenses.}
    \label{fig:pipeline}
\end{figure*}

So far, the raw images resulting from this camera array are only usable when looking at single images.
However, combining the images of multiple cameras yields unregistered results, i.e., when overlaying the images the same pixel shows the content of different objects.
An example for overlaying three raw images of HAHSI is shown in \fig\ref{fig:pipeline} in the bottom left.
That is due to their different spatial positions within the camera array, but also due to manufacturing deviations between different sensors and different lenses.
Moreover, different bandpass filters sitting in front of the lenses also influence the optical path to the sensor differently.
Finally, the camera-lens-filter combinations may be not perfectly aligned within the camera array as well.
To compensate all of these effects, an image processing pipeline is necessary, which is shown in \fig\ref{fig:pipeline}.
The goal of this pipeline is to map every camera to the center view.
The registration process needed to account for the spatial displacement between the cameras is the most elaborate part of this pipeline, since the other effects can be tackled using classical calibration procedures.
The registration algorithm is largely based on~\cite{msir_tip_2024}, where a registration process for a rectangular multispectral camera array with nine cameras is described.
Since the fundamental properties of this multispectral camera array and the proposed hyperspectral camera array are shared, i.e., different cameras record different spectral areas, these algorithms can be leveraged for HAHSI as well.
This registration pipeline is reviewed briefly in the following.

\subsection{Calibration}

As already mentioned, diverse manufacturing deviations are negatively influencing the assumption that the images are perfectly aligned according to the hexagonal shape of HAHSI.
Thus, an intrinsic calibration procedure to compensate the lens distortions is required.
This can be achieved by recording a well-defined pattern, for example a checkerboard pattern, multiple times at different positions of the images.
The models from Strobl and Hirzinger~\cite{strobl_calib_2011} are used.

An extrinsic camera calibration procedure is required for inter camera calibration.
Inter camera calibration has the goal of aligning the images according to the positions within the camera array.
Consequently, the epipolar constraint is fulfilled, which reduces the dimensionality of searching the same pixel in different views from a two-dimensional problem to a one-dimensional one.
This procedure also uses a checkerboard pattern and is described in detail in~\cite{genser_camsi_2020}.
The calibration procedure results in the calibrated image $\calibrated_\channel[\calcoord]$ for every channel $\channel$.

\subsection{Cross Spectral Disparity Estimation}

\begin{figure*}[t]
    \centering
    \input{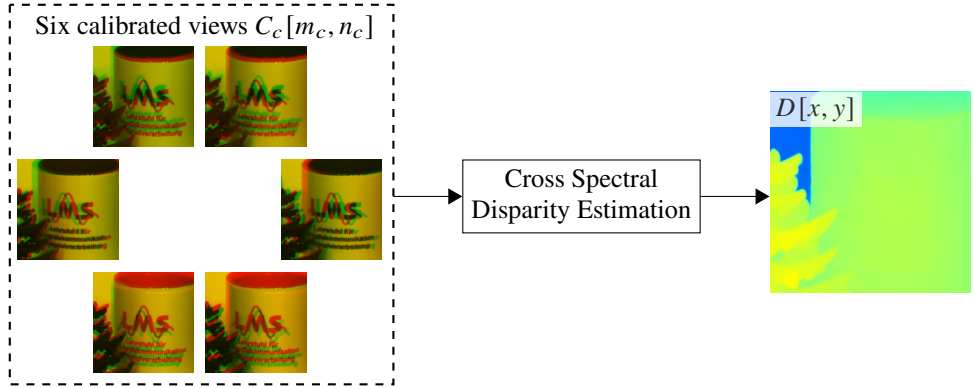}
    \caption{Six false color images are shown on the left. These show the center camera in the red channel and one peripheral cameras in the green channel. On the right, the fused estimated disparity map is depicted. Note that more peripheral cameras are used to estimate disparity during the registration process.}
    \label{fig:disparity_estimation}
\end{figure*}

Starting from this subsection, the images are assumed to be perfectly aligned and free of radial and tangential distortions.
In the registered hyperspectral image, the pixels of the same object shall be warped to the same position as in the center view.
Therefore, it is necessary to find the same object pixels in all images.
For that, disparity estimation for stereo cameras can be employed~\cite{msir_tip_2024}.
Disparity is the pixel distance between the same object pixel of two different views.
As shown in \fig\ref{fig:disparity_estimation}, the result of disparity estimation is a map $\disparity[\descoord]$, which provides this distance for every pixel of the center view.

\subsection{Warping and Occlusion Detection}

\begin{figure*}[t]
    \centering
    \input{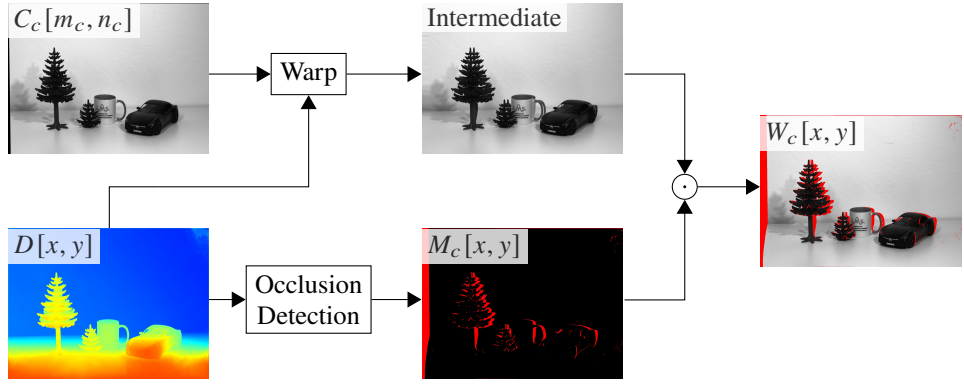}
    \caption{A calibrated image is warped to the center camera using the estimated disparity map. This disparity map is also used to detect occlusions. Finally, the occlusion map is multiplied to the warped image.}
    \label{fig:warp_and_occlusion}
\end{figure*}

We use the estimated disparity map $\disparity[\descoord]$ to warp the peripheral views to the center view, which results in the warped hyperspectral image $\hsimagewarp_\channel[\descoord]$ for channel $\channel$.
For this, one needs to account for the rotation and baseline of the peripheral cameras to the center camera.
As shown in \fig\ref{fig:warp_and_occlusion}, the problem with warping the peripheral views using the center disparity is that in occluded regions the occluding objects will be repeated in the background.
These regions correspond to the pixels which are occluded to the peripheral views.
Hence, an occlusion detection algorithm~\cite{msir_tip_2024} is necessary to estimate an occlusion map $\mask_\channel[\descoord]$ to mask these regions.

\subsection{Cross Spectral Reconstruction}

\begin{figure}[t]
    \centering
    \input{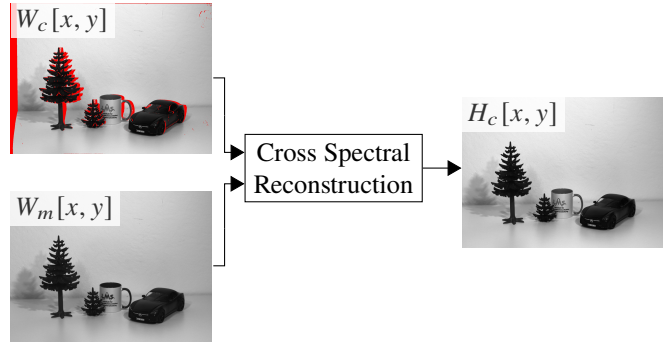}
    \caption{The cross spectral reconstruction module exploits the structure of the center camera to reconstruct the occluded pixels (red) of the peripheral camera.}
    \label{fig:reconstruction}
\end{figure}

The last step within the registration is to reconstruct the occluded pixels to produce the registered hyperspectral image $\hsimage_\channel[\descoord]$.
Fortunately, as shown in \fig\ref{fig:reconstruction}, the image of the center camera $\hsimagewarp_\centercam[\descoord]$ is fully available and can serve as guide.
Therefore, the goal is to exploit the structure of the guide to reconstruct the missing pixels.
For this, a deep guided neural network~\cite{msir_tip_2024} is employed, which works using linear regression coefficients, where the center image serves as guide image.

\section{Synthetic HAHSI Data}
\label{sec:synth_data}
\begin{figure}[t]
    \centering
    \input{tikz/synth_data.tikz}
    \caption{Sample images from the synthetic HAHSI dataset. All seven synthetic scenes are depicted as RGB image, which was rendered using simulated color curves on all 37 hyperspectral channels.}
    \label{fig:synth_data}
\end{figure}

The evaluation of the performance of this hyperspectral camera array cannot be done on real-world data, since it is not possible to record ground-truth data with this setup.
For real ground-truth data, one first has to capture a scene using every bandpass filter in front of the center camera.
Then, the problem is that these filters slightly change the optical path of the light.
Hence, a registration procedure would still be necessary and this data cannot be called ground-truth data anymore.
Therefore, a synthetic hyperspectral database is introduced, where the scenes are rendered in Blender~\cite{blender} using a digital twin of the real-world camera array.
Blender can render realistic images using its branched path tracing algorithm.
The basis of this principle is the synthetic hyperspectral database provided in~\cite{sippel_hyvid_2023}.
However, this database was only rendered using a three by three camera array.
For this paper, we render this database using our hexagonal design as well as an orthogonal-spaced version with the same baseline between cameras for comparison.
Moreover, the spectral area in this database ranged from 400 nm to 700 nm, which was extended to 760 nm to cover the proposed camera array as well.
The scenes are rendered using the specified spectral band except the center camera, for which all bands are rendered to generate the ground-truth data.
This process speeds up rendering time significantly.
The database is available at \url{https://github.com/FAU-LMS/HAHSI}.

\subsection{Textures}

The textures for the database are extracted using another real-world hyperspectral database~\cite{arad_bgu_2016}.
This database contains over 200 hyperspectral images from different scenarios including urban, indoor and outdoor scenes.
Thus, this database is suitable to extract hyperspectral images for diverse scenarios.
The images itself are extracted my manually selecting four points such that the area includes the texture to extract.
Then, a homography matrix is calculated from these four points.
With this homography matrix the texture can be warped to a proper two-dimensional space.
This procedure is described in detail in~\cite{sippel_hyvid_2023}.

\subsection{Light}

The light sources are emulated using Planck's law.
For outdoor scenes, the sun is using a color temperature of 6400K, which is the temperature that the sun reaches at bright daylight.
For indoor scenes, the lamps are emulating a light source with color temperature of 3200K.

\subsection{Scenes}

In total, seven scenes with movement were rendered for 30 frames each.
The scenes are \textit{family house}, \textit{medieval seaport}, \textit{city}, \textit{outdoor}, \textit{forest}, \textit{indoor} and \textit{lab}.
While the \textit{indoor} scene still uses the sun as light source shining through the windows, lamps are used for the \textit{lab} scene.
In all scenes, the camera array is moving through the scene.
Moreover, in scene \textit{city} a car is moving on the street.
One frame of all scenes is depicted in \fig\ref{fig:synth_data}.

\section{Evaluation}
\label{sec:evaluation}
In this section, the synthetic data introduced in the previous section is used to quantitatively evaluate the proposed array against its main snapshot competitors, namely filter arrays and coded apertures.

\subsection{Filter Array Simulation}

\begin{figure}[t]
    \centering
    \input{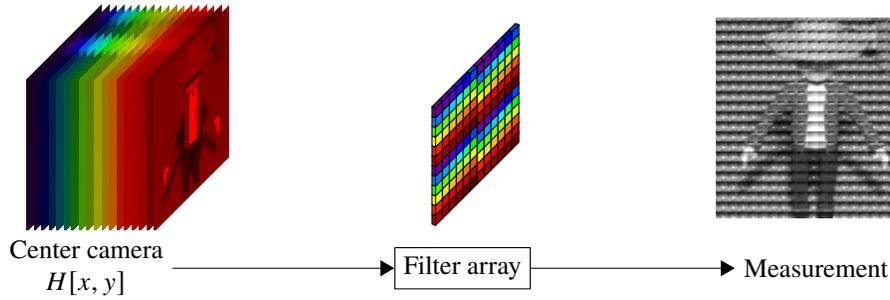}
    \caption{The image processing required to simulate a hyperspectral filter array using the center camera of the synthetic database.}
    \label{fig:eval_msfa}
\end{figure}

As shown in \fig\ref{fig:eval_msfa}, filter arrays~\cite{monno_msfa_2015} can be easily simulated on the proposed synthetic data, since the hyperspectral image just needs to be periodically sampled at the corresponding wavelengths.
For that, the hyperspectral image of the center camera is used.
The measurement is created by applying the periodic hyperspectral filter array to this hyperspectral image.
To simulate the filtering process, the channel matching the corresponding filter is used as grayscale value for this pixel.
Unfortunately, 37 is a prime number and thus cannot be split into a rectangle.
Therefore, the evaluation is based on the first 36 channels, such that a 6 \texttimes\ 6 filter array can be simulated.
Different reconstruction methods for multispectral filter arrays are evaluated, namely, wavelet-based multispectral demosaicing (DWT)~\cite{dwt_2016}, weighted bilinear interpolation (WBI)~\cite{spectral_differences_2006}, spectral differences (SD)~\cite{spectral_differences_2006} and iterative spectral differences (ISD)~\cite{iterative_spectral_differences_2014}.

\subsection{Coded Aperture Simulation}

\begin{figure}[t]
    \centering
    \input{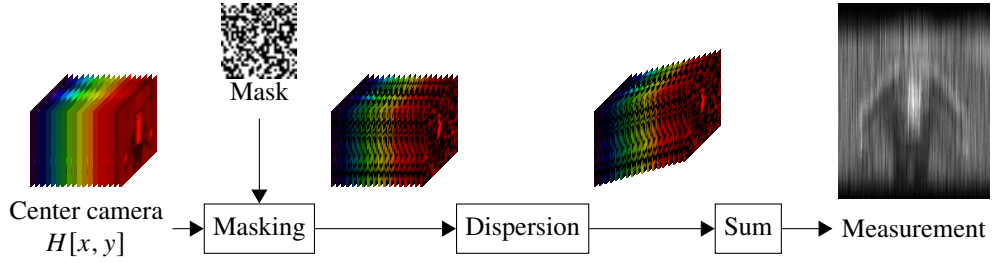}
    \caption{The image processing required to simulate a coded aperture approach using the center camera of the synthetic database.}
    \label{fig:eval_cassi}
\end{figure}

As shown in \fig\ref{fig:eval_cassi}, coded aperture snapshot spectral imaging (CASSI)~\cite{gehm_cassi_2007} can be simulated using this synthetic database as well.
The hyperspectral image of the center camera is used to produce the measurement of CASSI.
First, the mask necessary to code the aperture is generated using blue noise~\cite{blue_noise_2016}.
Then, this mask is multiplied to the hyperspectral image of the center camera.
Afterwards, the dispersive element is simulated by translating the channel image of each wavelength step, here 10 nm, by one pixel in comparison to the previous wavelength step.
This assumption is better than a dispersive element would perform in real life, which is advantageous for the compressive sensing task and therefore positively influences the evaluation metrics.
Finally, this coded and translated hyperspectral datacube is summed up to simulate the sensor and produce the measurement.

Note that in literature often several records with different masks are used to reconstruct one scene.
However, here snapshot capable cameras are evaluated and thus, the simulated system is only allowed to record the scene at one timestep using one mask.
The images of the simulated CASSI are reconstructed by generalized alternating projection based total variation minimization for compressive sensing (GAP)~\cite{cassi_gap_2016}.

\subsection{Quantitative Results}

The different techniques are compared to each other by comparing the reconstructed hyperspectral image with the ground truth hyperspectral image using the Peak Signal-to-Noise Ratio (PSNR) and the Structural Similarity Index (SSIM).
The PSNR is defined by the Mean Squared Error (MSE)
\begin{equation}
    \text{MSE}(\hsimage, \groundtruth) = \frac{1}{\imagewidth \imageheight \numchannels} \sum_{\descoordX=0}^{\imagewidth - 1} \sum_{\descoordY=0}^{\imageheight - 1} \sum_{c=0}^{\numchannels - 1} (\hsimage_c[\descoord] - \groundtruth_c[\descoord])^2,
\end{equation}
where $\hsimage[\descoord]$ is a hyperspectral image with $\numchannels$ channels of size $\imagewidth \times \imageheight \times \numchannels$ and $\groundtruth[\descoord]$ its corresponding ground truth.
Hence, the PSNR can be calculated by
\begin{equation}
    \text{PSNR}(\hsimage, \groundtruth) = 20 \log_{10}\left( \hsimage_{\max} \right) - 10 \log_{10}(\text{MSE}(\hsimage, \groundtruth)),
\end{equation}
where $\hsimage_{\max}$ is the peak signal value of the hyperspectral image.
Originally, the images were are rendered and recorded using 8 bits, thus, $\hsimage_{\max}$ corresponds to 255.
However, the images are normalized to the range of $[0, 1]$ and all the processing happens in this value range.
Therefore, for the upcoming evaluation $\hsimage_{\max}$ is set to 1.

The second evaluation metric is SSIM, which aims at better correlating with human perception.
For this, luminance, contrast and structure between windows of the registered image and the ground truth image are combined in one formula based on first-order and second-order statistical moments.
The combination leads to~\cite{ssim_2003}
\begin{equation}
    \text{SSIM}(\hsimage, \groundtruth) = \frac{\left( 2 \mean_\hsimage\mean_\groundtruth + \ssimconst_1 \right) \left( 2 \variance_{\hsimage\groundtruth} + \ssimconst_2 \right)}{\left(\mean_\hsimage^2 + \mean_\groundtruth^2 + \ssimconst_1 \right)\left(\variance_\hsimage^2 + \variance_\groundtruth^2 + \ssimconst_2 \right)},
\end{equation}
where $\mean_\hsimage$ and $\mean_\groundtruth$ is the mean of windows of $\hsimage$ and $\groundtruth$, respectively.
Similarly, $\variance_\hsimage$ and $\variance_\groundtruth$ are the corresponding variances and $\variance_{\hsimage\groundtruth}$ is the covariance between $\hsimage$ and $\groundtruth$.
$\ssimconst_1$ and $\ssimconst_2$ are constants that account for small denominators and hence stability issues.
For the final SSIM value, all windows are averaged.

\begin{figure*}
    \centering
    \begin{tikzpicture}[>=triangle 60, y=-1cm]
    \begin{scope} [scale=0.75, shift={(0, 0)}]
        \draw[line width=1] (-0.5, 0.5) to[out=45,in=135, distance=0.2cm] (0.5, 0.5);
        \draw[line width=1] (-0.5, 0.5) to[out=-45,in=-135, distance=0.2cm] (0.5, 0.5);

        \draw[line width=1] (-0.5, -0.5) to[out=45,in=135, distance=0.2cm] (0.5, -0.5);
        \draw[line width=1] (-0.5, -0.5) to[out=-45,in=-135, distance=0.2cm] (0.5, -0.5);

        \draw[line width=1] (-0.5, -0.5) to (-0.5, 0.5);
        \draw[line width=1] (0.5, -0.5) to (0.5, 0.5);

        \node[align=center] (db) at (0, 0) {DB};
    \end{scope}

    \node[draw, align=center] (cc) at (0, 2) {Extract\\center view};

    \node[draw, align=center] (sh) at (4, -2) {Simulate\\HAHSI};
    \node[draw, align=center] (sf) at (4, 0) {Simulate\\filter array};
    \node[draw, align=center] (sc) at (4, 2) {Simulate\\CASSI};

    \node[draw, align=center] (reg) at (8, -2) {Registration};
    \node[draw, align=center] (dem) at (8, 0) {Demosaicing};
    \node[draw, align=center] (rec) at (8, 2) {Reconstruction};

    \node[draw, align=center] (el1) at (11.5, -2) {Evaluation};
    \node[draw, align=center] (el2) at (12, 0) {Evaluation};
    \node[draw, align=center] (el3) at (12.5, 2) {Evaluation};

    \node[align=center] (ps1) at (15.5, -2) {PSNR\\SSIM};
    \node[align=center] (ps2) at (15.5, 0) {PSNR\\SSIM};
    \node[align=center] (ps3) at (15.5, 2) {PSNR\\SSIM};

    \draw[->] (0, -0.45) |- (sh.west);
    \draw[->] (0, 0.45) -- (cc.north);

    \draw[->] (cc.east) -- (sc.west);
    \draw[->] (1.5, 2) |- (sf.west);
    \draw[fill=black] (1.5, 2) circle (0.5mm);

    \draw[->] (sh.east) -- (reg.west);
    \draw[->] (sf.east) -- (dem.west);
    \draw[->] (sc.east) -- (rec.west);

    \draw[->] (reg.east) -- (el1.west);
    \draw[->] (dem.east) -- (el2.west);
    \draw[->] (rec.east) -- (el3.west);

    \draw[->] (el1.east) -- (ps1.west);
    \draw[->] (el2.east) -- (ps2.west);
    \draw[->] (el3.east) -- (ps3.west);

    \draw[-] (cc.south) |- (10.8, 3);
    \draw[fill=black] (10.8, 3) circle (0.5mm);
    \draw[-] (10.8, 3) -- (10.8, 2.1);
    \draw[-] (10.8, 1.9) -- (10.8, 0.1);
    \draw[->] (10.8, -0.1) -- (10.8, -1.75);

    \draw[-] (10.8, 3) -| (11.3, 2.1);
    \draw[fill=black] (11.3, 3) circle (0.5mm);
    \draw[->] (11.3, 1.9) -- (11.3, 0.25);

    \draw[->] (10.8, 3) -| (11.8, 2.25);
\end{tikzpicture}
    \caption{A signal flow diagram of the evaluation and how PSNR and SSIM values are produced using the synthetic database as starting point. Note that the filter array and CASSI only use the center camera of the synthetic database.}
    \label{fig:evaluation}
\end{figure*}
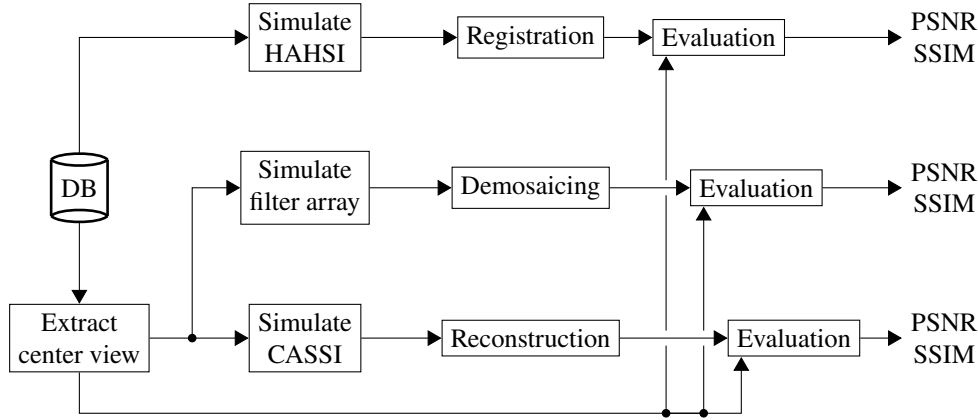

The signal flow to compare HAHSI to hyperspectral filter arrays and CASSI is shown in \fig\ref{fig:evaluation}.
The basis for the evaluation is the synthetic hyperspectral array database.
With this database, HAHSI can be easily simulated by selecting the corresponding spectral band for each camera.
Filter arrays and CASSI only need one camera.
Hence, the center camera is extracted from the database and used to simulate both approaches.
Afterwards, the registration, demosaicing or reconstruction algorithms are used to produce consistent hyperspectral data cubes, respectively.
Finally, these reconstructed hyperspectral data cubes can be compared to the ground truth, namely, the hyperspectral data cube of the center cam, to calculate PSNR and SSIM values.

\begin{table}[t]
	\footnotesize
	\centering
	\caption{Evaluation of the proposed snapshot hyperspectral imaging system against its snapshot competitors on full images. The results are given by PSNR in dB / SSIM.}
	\label{tab:eval}
	\begin{tabular}{@{\hspace*{0.0cm}}l@{\hspace*{0.1cm}}|@{\hspace*{0.1cm}}c@{\hspace*{0.1cm}}|@{\hspace*{0.1cm}}c@{\hspace*{0.1cm}}c@{\hspace*{0.1cm}}c@{\hspace*{0.1cm}}c@{\hspace*{0.1cm}}|@{\hspace*{0.1cm}}c@{\hspace*{0.0cm}}}
        & CASSI & \multicolumn{4}{c|@{\hspace*{0.1cm}}}{MSFA} & HAHSI \\
        & GAP~\cite{cassi_gap_2016} & DWT~\cite{dwt_2016} & WBI~\cite{spectral_differences_2006} & SD~\cite{spectral_differences_2006} & ISD~\cite{iterative_spectral_differences_2014} & Proposed \\
        \hline
        F. house    & 19.64/.726 & 22.80/.735 & 23.12/.778  & 25.71/.849 & 26.45/.852 & \textbf{29.42}/\textbf{.962} \\
        Med. sea.   & 22.45/.762 & 32.50/.859 & 32.89/.877  & 35.21/.930 & 35.80/.931 & \textbf{43.95}/\textbf{.988} \\
        City        & 23.65/.834 & 31.52/.901 & 31.86/.917  & 34.26/.946 & 35.00/.946 & \textbf{40.27}/\textbf{.988} \\
        Outdoor     & 19.69/.660 & 22.77/.640 & 22.99/.675  & 25.44/.808 & 26.24/.822 & \textbf{28.82}/\textbf{.917} \\
        Forest      & 21.58/.691 & 24.07/.671 & 24.27/.716  & 26.85/.804 & \textbf{27.68}/.821 & 27.67/\textbf{.892} \\
        Indoor      & 23.53/.824 & 32.76/.936 & 33.50/.953  & 35.55/.958 & \textbf{36.09}/.957 & 35.17/\textbf{.983} \\
        Lab         & 18.84/.647 & 32.45/.953 & 34.50/\textbf{.973}  & 35.13/.966 & \textbf{35.37}/.964 & 28.51/.969 \\
        \hline
        Average     & 21.34/.735 & 28.41/.813 & 29.02/.841  & 31.17/.894 & 31.80/.899 & \textbf{33.40}/\textbf{.957} \\
	%\vspace*{0.2cm}
	\end{tabular}
\end{table}

\begin{table}[t]
	\footnotesize
	\centering
	\caption{Evaluation of the proposed snapshot hyperspectral imaging system against its snapshot competitors on the intersected image area. The results are given by PSNR in dB / SSIM.}
	\label{tab:eval_intersection}
	\begin{tabular}{@{\hspace*{0.0cm}}l@{\hspace*{0.1cm}}|@{\hspace*{0.1cm}}c@{\hspace*{0.1cm}}|@{\hspace*{0.1cm}}c@{\hspace*{0.1cm}}c@{\hspace*{0.1cm}}c@{\hspace*{0.1cm}}c@{\hspace*{0.1cm}}|@{\hspace*{0.1cm}}c@{\hspace*{0.0cm}}}
        & CASSI & \multicolumn{4}{c|@{\hspace*{0.1cm}}}{MSFA} & HAHSI \\
        & GAP~\cite{cassi_gap_2016} & DWT~\cite{dwt_2016} & WBI~\cite{spectral_differences_2006} & SD~\cite{spectral_differences_2006} & ISD~\cite{iterative_spectral_differences_2014} & Proposed \\
        \hline
        F. house    & 19.44/.720 & 22.26/.707 & 22.58/.756 & 25.18/.834 & 25.94/.840 & \textbf{29.00}/\textbf{.959} \\
        Med. sea.   & 22.73/.768 & 32.66/.864 & 33.07/.883 & 35.33/.931 & 35.90/.932 & \textbf{45.15}/\textbf{.989} \\
        City        & 23.35/.826 & 31.42/.899 & 31.83/.917 & 34.22/.943 & 34.92/.943 & \textbf{41.20}/\textbf{.988} \\
        Outdoor     & 19.55/.652 & 22.15/.606 & 22.37/.643 & 24.83/.791 & 25.64/.807 & \textbf{28.59}/\textbf{.909} \\
        Forest      & 21.73/.680 & 24.06/.654 & 24.24/.699 & 26.81/.796 & 27.65/.815 & \textbf{28.61}/\textbf{.894} \\
        Indoor      & 23.73/.829 & 32.29/.928 & 32.94/.947 & 35.02/.954 & 35.56/.953 & \textbf{39.10}/\textbf{.984} \\
        Lab         & 18.68/.644 & 31.31/.939 & 33.38/.966 & 33.99/.956 & \textbf{34.21}/.954 & 29.63/\textbf{.972} \\
        \hline
        Average     & 21.32/.731 & 28.02/.800 & 28.63/.830 & 30.77/.886 & 31.40/.892 & \textbf{34.47}/\textbf{.956} \\
	%\vspace*{0.2cm}
	\end{tabular}
    \vspace*{0.3cm}
\end{table}

The results are summarized in \tab\ref{tab:eval} and \tab\ref{tab:eval_intersection}.
\tab\ref{tab:eval} shows the results on full $2448\times2048$ images, which is disadvantageous for our proposed method.
Due to the spatial distribution of the cameras and the large baseline in comparison to the simulated working distance, the peripheral cameras are not able to capture all border pixels of the center camera, depending on their position within the array.
Hence, a lot of border pixels need to be reconstructed.
Therefore, the evaluation is also done on the intersected area of the images, where 200 pixels on each border of the images are not considered during evaluation, which results in $2048 \times 1648$ pixel hyperspectral images.
Of course, the intersected area depends on the minimum depth recorded by the camera array.
The closer objects appear in the camera, the smaller the intersected area.
For the synthetic database, 200 pixels is a good approximation for this circumstance.
These results are shown in \tab\ref{tab:eval_intersection}.

HAHSI clearly outperforms all of its competitors on average over all scenes on full images, while this gain increases even more when considering the intersected area.
Interestingly, when evaluating full images of the scenes \textit{forest}, \textit{indoor} and \textit{lab} in terms of PSNR, HAHSI performs worse than an MSFA with a good reconstruction algorithm.
While \textit{forest} is a scene with a lot of high frequency content, \textit{indoor} and \textit{lab} also contain a lot of low frequency content.
This is closely connected to the performance of the disparity estimation, which explains the performance of HAHSI for scene \textit{lab}.
For this scene, the disparity estimation in the uniform area is completely off, which then leads to many occluded pixels that need to be reconstructed.
The cross spectral reconstruction module fails to properly reconstruct some of these pixels in some channels.
Nevertheless, HAHSI still outperforms MSFAs by more than 1.5 dB in terms of PSNR and a much better SSIM on full images on average.
Moreover, a total gain of 3 dB in terms of PSNR is achieved when considering the intersected area.
A similar picture is drawn when comparing SSIM values, where the gain over the MSFAs with ISD increases as well.
While HAHSI is getting even better in terms of PSNR when considering the intersected area, the other algorithms are slightly worse.
Moreover, one-shot CASSI is not able to compete with MSFAs and HAHSI by a significant margin.
Using more measurements, CASSI could get better, but looses snapshot capability.

\subsection{Qualitative Analysis}

\begin{figure}
    \centering
    \input{tikz/eval_qualitative.tikz}
    \caption{Qualitative evaluation of channel 10 of the first frame of the synthetic scene \textit{family house}.}
    \label{fig:eval_qualitative}
\end{figure}

In \fig\ref{fig:eval_qualitative} the ground truth and reconstructed images of channel 10 of a single frame of the scene \textit{family house} is depicted.
Single-shot CASSI is not able to keep up with HAHSI and the reconstruction algorithms for MSFA.
Moreover, it is visible that the reconstruction algorithms for MSFA get increasingly sharper with ISD being the best method.
However, in comparison to HAHSI, ISD still looks much more blurry.
In general, the qualitative results align well with the quantitative results.

\subsection{Spectral Evaluation}

\begin{table}[t]
	\footnotesize
	\centering
	\caption{Evaluation of the proposed snapshot hyperspectral imaging system against its snapshot competitors on the intersected image area. The results are given in terms of spectral angle.}
	\label{tab:eval_intersection_sa}
	\begin{tabular}{@{\hspace*{0.0cm}}l@{\hspace*{0.1cm}}|@{\hspace*{0.1cm}}c@{\hspace*{0.1cm}}|@{\hspace*{0.1cm}}c@{\hspace*{0.1cm}}c@{\hspace*{0.1cm}}c@{\hspace*{0.1cm}}c@{\hspace*{0.1cm}}|@{\hspace*{0.1cm}}c@{\hspace*{0.0cm}}}
        & CASSI & \multicolumn{4}{c|@{\hspace*{0.1cm}}}{MSFA} & HAHSI \\
        & GAP~\cite{cassi_gap_2016} & DWT~\cite{dwt_2016} & WBI~\cite{spectral_differences_2006} & SD~\cite{spectral_differences_2006} & ISD~\cite{iterative_spectral_differences_2014} & Proposed \\
        \hline
        F. house    & 0.240 & 0.116 & 0.110 & 0.134 & 0.127 & \textbf{0.067} \\
        Med. sea.   & 0.216 & 0.031 & 0.031 & 0.034 & 0.033 & \textbf{0.011} \\
        City        & 0.180 & 0.031 & 0.030 & 0.035 & 0.034 & \textbf{0.017} \\
        Outdoor     & 0.265 & 0.164 & 0.168 & 0.203 & 0.188 & \textbf{0.117} \\
        Forest      & 0.372 & 0.249 & \textbf{0.243} & 0.301 & 0.285 & 0.254 \\
        Indoor      & 0.205 & 0.032 & 0.028 & 0.032 & 0.031 & \textbf{0.026} \\
        Lab         & 0.264 & 0.019 & \textbf{0.014} & 0.017 & 0.017 & 0.041 \\
        \hline
        Average     & 0.249 & 0.092 & 0.089 & 0.108 & 0.102 & \textbf{0.076} \\
	%\vspace*{0.2cm}
	\end{tabular}
    \vspace*{0.3cm}
\end{table}

Not only metrics which are tailored towards spatial image quality are crucial to evaluate the performance of different approaches, but also the performance regarding spectral accuracy is important.
Thus, the different approaches are also evaluated using the spectral angle between the reconstructed spectra and the ground truth spectra.
The spectral angle is calculated by
\begin{equation}
    \theta(\spectrum, \hat{\spectrum}) = \arccos(\frac{\spectrum\T}{||\spectrum||_2} \frac{\hat{\spectrum}\T}{||\hat{\spectrum}||_2})
\end{equation}
where $\spectrum$ is the ground-truth spectrum of a single pixel and $\hat{\spectrum}$ is the corresponding estimated spectrum.
A lower spectral angle denotes a better estimation, while the worst result an estimated spectrum can have is $\pi$.
For the evaluation, the spectral angles are averaged over whole hyperspectral images.
Ground-truth pixels with an all-zeros spectrum do not contribute to this average, while wrong all-zero estimations result in a spectral angle of $\pi$.

In \tab\ref{tab:eval_intersection_sa}, the quantitative results in terms of spectral angle are shown.
Again, HAHSI outperforms all other methods in terms of spectral angle and single-shot CASSI is not able to compete with all other methods.
Interestingly, in contrast to the evaluation on spatial image metrics, WBI outperforms the other demosaicing methods in terms of spectral angle.

\begin{figure}[t]
    \centering
    \input{images/spectra/plot.pgf}
    \caption{Qualitative evaluation of two spectra. These spectra belong to the synthetic scene \textit{forest}.}
    \label{fig:eval_spectra}
\end{figure}

In \fig\ref{fig:eval_spectra}, estimated spectra and the ground truth of two different exemplary pixels are depicted.
The spectrum on the left is accurately reconstructed by nearly all methods except CASSI, which is only able to reconstruct the rough shape of the ground truth.
On the other hand, the spectrum on the right is only accurately estimated by HAHSI.
The results of SD and ISD are also usable to some extent, while the results of DWT, WBI and CASSI are off.
In general, these two plots support the quantative results of \tab\ref{tab:eval_intersection_sa}.

\subsection{Ground-truth Disparity Evaluation}
\label{sec:eval_disp_eval}

\begin{table}[t]
	\footnotesize
	\centering
	\caption{Evaluation of using ground-truth disparity maps in comparison to the estimated disparity on the synthetic HAHSI data on full images and considering the intersected area. The results are given by PSNR in dB / SSIM.}
	\label{tab:eval_gt_depth}
        \begin{tabular}{@{\hspace*{0.0cm}}l@{\hspace*{0.1cm}}|@{\hspace*{0.1cm}}c@{\hspace*{0.1cm}}c@{\hspace*{0.1cm}}}
        \textbf{Full}& Estimated   & Ground truth   \\
        \hline
        Family h.   & 29.42/0.962 & \textbf{30.95}/\textbf{0.970}  \\
        Med. sea.   & 43.95/\textbf{0.988} & \textbf{44.26}/\textbf{0.988}  \\
        City        & 40.27/0.988 & \textbf{41.29}/\textbf{0.989}  \\
        Outdoor     & 28.82/0.917 & \textbf{30.38}/\textbf{0.949}  \\
        Forest      & \textbf{27.67}/0.892 & 27.33/\textbf{0.914}  \\
        Indoor      & 35.17/0.983 & \textbf{35.20}/\textbf{0.986}  \\
        Lab         & 28.51/0.969 & \textbf{31.14}/\textbf{0.977}  \\
        \hline
        Average     & 33.40/0.957 & \textbf{34.36}/\textbf{0.968}  \\
    \end{tabular}
    \begin{tabular}{@{\hspace*{0.1cm}}l@{\hspace*{0.1cm}}|@{\hspace*{0.1cm}}c@{\hspace*{0.1cm}}c@{\hspace*{0.0cm}}}
        \textbf{Intersection}& Estimated   & Ground truth   \\
        \hline
        Family h.   & 29.00/0.959 & \textbf{30.53}/\textbf{0.968}  \\
        Med. sea.   & 45.15/\textbf{0.989} & \textbf{45.68}/\textbf{0.989}  \\
        City        & 41.20/0.988 & \textbf{42.39}/\textbf{0.989}  \\
        Outdoor     & 28.59/0.909 & \textbf{30.78}/\textbf{0.948}  \\
        Forest      & 28.61/0.894 & \textbf{29.43}/\textbf{0.922}  \\
        Indoor      & 39.10/0.984 & \textbf{41.34}/\textbf{0.987}  \\
        Lab         & 29.63/0.972 & \textbf{33.66}/\textbf{0.980}  \\
        \hline
        Average     & 34.47/0.956 & \textbf{36.26}/\textbf{0.969}  \\
    \end{tabular}
\end{table}

The quality of the resulting hyperspectral images and videos heavily depends on the registration algorithms.
Hence, better registration procedures also lead to better hyperspectral images and videos.
To see this, \tab\ref{tab:eval_gt_depth} shows PSNR and SSIM results on the synthetic data reconstructed using the ground-truth disparity in comparison to the estimated disparity.
This table shows the potential quality increases if the disparity estimation gets better.
Depending on the scene, improvements of more than 2 dB for the full image evaluation and more than 4 dB when considering the intersected area.
On average, the performance can be improved by nearly 1 dB in case of full images and nearly 2 dB when considering the intersected area.
In summary, this evaluation showed that the performance advantage of HAHSI shown in \tab\ref{tab:eval} and \tab\ref{tab:eval_intersection} can be further increased by developing further improved registration procedures.

\subsection{Camera Array Shape Evaluation}
\label{subsec:shape_eval}

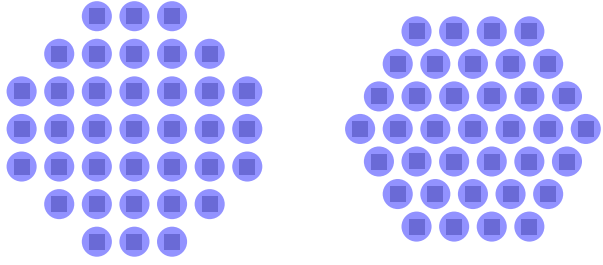
\begin{figure}[t]
    \centering
    \begin{tikzpicture}[scale=0.5, >=triangle 60]
    \definecolor{camerablue}{RGB}{100,100,255}
    \definecolor{cameragray}{RGB}{120,120,120}
    \def\sensorsize{0.4}
    \def\lenssize{0.4}
    \begin{scope} [shift={(-3, -3)}]
        \foreach \x in {0,...,6}
        {
            \foreach \y in {0,...,6}
            {
                \pgfmathparse{(\x - 3) * (\x - 3) + (\y - 3) * (\y - 3) - 3.5 * 3.5 <= 0 ? int(-1) : int(1)}
                \ifnum \pgfmathresult < 0
                \fill[cameragray] (\x - \sensorsize/2, \y - \sensorsize/2) rectangle (\x + \sensorsize/2, \y + \sensorsize/2);
                \fill[camerablue, opacity=0.7] (\x, \y) circle (\lenssize);
                \fi
            }
        }
        \centerarc[red,thick](6,4)(0:45:\lenssize)
        \connectionline[red,thick](6,4)(4,6)(45:\lenssize)
        \centerarc[red,thick](4,6)(45:90:\lenssize)
        \connectionline[red,thick](4,6)(2,6)(90:\lenssize)
        \centerarc[red,thick](2,6)(90:135:\lenssize)
        \connectionline[red,thick](2,6)(0,4)(135:\lenssize)
        \centerarc[red,thick](0,4)(135:180:\lenssize)
        \connectionline[red,thick](0,4)(0,2)(180:\lenssize)
        \centerarc[red,thick](0,2)(180:225:\lenssize)
        \connectionline[red,thick](0,2)(2,0)(225:\lenssize)
        \centerarc[red,thick](2,0)(225:270:\lenssize)
        \connectionline[red,thick](2,0)(4,0)(270:\lenssize)
        \centerarc[red,thick](4,0)(270:315:\lenssize)
        \connectionline[red,thick](4,0)(6,2)(315:\lenssize)
        \centerarc[red,thick](6,2)(315:360:\lenssize)
        \connectionline[red,thick](6,2)(6,4)(0:\lenssize)
        \node[red] at (3, 6.85) {\SI{116613}{\milli\metre\squared}};
        \draw[->, lightgray] (6.6, 3) -- (7.9, 3);
    \end{scope}
    \def\inner{6}
    \def\innerradius{1}
    \def\middle{12}
    \def\middleradiussmall{1.7320508075688772}
    \def\middleradiusbig{2}
    \def\outer{18}
    \def\outerradiussmall{2.6457513110645907}
    \def\outerradiusbig{3}
    \def\angleone{19.106605350869096}
    \def\angletwo{40.893394649130904}
    \def\anglethree{60}
    \begin{scope} [shift={(8.5, 0)}]
        % rect shape for comparison
        \begin{scope} [shift={(-3, -3)}]
            \centerarc[lightgray,thick,opacity=1.0](6,4)(0:45:\lenssize)
            \connectionline[lightgray,thick,opacity=1.0](6,4)(4,6)(45:\lenssize)
            \centerarc[lightgray,thick,opacity=1.0](4,6)(45:90:\lenssize)
            \connectionline[lightgray,thick,opacity=1.0](4,6)(2,6)(90:\lenssize)
            \centerarc[lightgray,thick,opacity=1.0](2,6)(90:135:\lenssize)
            \connectionline[lightgray,thick,opacity=1.0](2,6)(0,4)(135:\lenssize)
            \centerarc[lightgray,thick,opacity=1.0](0,4)(135:180:\lenssize)
            \connectionline[lightgray,thick,opacity=1.0](0,4)(0,2)(180:\lenssize)
            \centerarc[lightgray,thick,opacity=1.0](0,2)(180:225:\lenssize)
            \connectionline[lightgray,thick,opacity=1.0](0,2)(2,0)(225:\lenssize)
            \centerarc[lightgray,thick,opacity=1.0](2,0)(225:270:\lenssize)
            \connectionline[lightgray,thick,opacity=1.0](2,0)(4,0)(270:\lenssize)
            \centerarc[lightgray,thick,opacity=1.0](4,0)(270:315:\lenssize)
            \connectionline[lightgray,thick,opacity=1.0](4,0)(6,2)(315:\lenssize)
            \centerarc[lightgray,thick,opacity=1.0](6,2)(315:360:\lenssize)
            \connectionline[lightgray,thick,opacity=1.0](6,2)(6,4)(0:\lenssize)
        \end{scope}

        % center
        \fill[cameragray] (-\sensorsize/2, -\sensorsize/2) rectangle (\sensorsize/2, \sensorsize/2);
        \fill[camerablue, opacity=0.7] (0, 0) circle (\lenssize);

        \foreach \i in {1,...,\inner}
        {
            \fill[cameragray] ({\innerradius * cos((\i/\inner)*360) - \sensorsize/2}, {\innerradius * sin((\i/\inner)*360) - \sensorsize/2})
            rectangle ({\innerradius * cos((\i/\inner)*360) + \sensorsize/2}, {\innerradius * sin((\i/\inner)*360) + \sensorsize/2});
            \fill[camerablue, opacity=0.7] ({\innerradius * cos((\i/\inner)*360)}, {\innerradius * sin((\i/\inner)*360)}) circle (\lenssize);
        }

        \foreach \i in {1,...,\middle}
        {
            \pgfmathsetmacro{\size}{ifthenelse(Mod(\i,2) == 0, \middleradiusbig, \middleradiussmall)}
            \fill[cameragray] ({\size * cos((\i/\middle)*360) - \sensorsize/2}, {\size * sin((\i/\middle)*360) - \sensorsize/2})
                rectangle ({\size * cos((\i/\middle)*360) + \sensorsize/2}, {\size * sin((\i/\middle)*360) + \sensorsize/2});
            \fill[camerablue, opacity=0.7] ({\size * cos((\i/\middle)*360)}, {\size * sin((\i/\middle)*360)}) circle (\lenssize);
        }

        \foreach \i in {1,...,\outer}
        {
            \pgfmathsetmacro{\size}{ifthenelse(Mod(\i,3) == 0, \outerradiusbig, \outerradiussmall)}
            \pgfmathsetmacro{\angle}{ifthenelse(Mod(\i,3) == 0, \anglethree + \i * 20, ifthenelse(Mod(\i,3) == 1, \angleone + (\i - 1) * 20, \angletwo + (\i - 2) * 20)}
            \fill[cameragray] ({\size * cos(\angle) - \sensorsize/2}, {\size * sin(\angle) - \sensorsize/2})
                rectangle ({\size * cos(\angle) + \sensorsize/2}, {\size * sin(\angle) + \sensorsize/2});
            \fill[camerablue, opacity=0.7] ({\size * cos(\angle)}, {\size * sin(\angle)}) circle (\lenssize);
        }

        \centerarc[red,thick](1.5,3 * 0.866)(30:90:\lenssize)
        \connectionline[red,thick](1.5,3 * 0.866)(-1.5,3 * 0.866)(90:\lenssize)
        \centerarc[red,thick](-1.5,3 * 0.866)(90:150:\lenssize)
        \connectionline[red,thick](-1.5,3 * 0.866)(-3,0)(150:\lenssize)
        \centerarc[red,thick](-3,0)(150:210:\lenssize)
        \connectionline[red,thick](-3,0)(-1.5,-3 * 0.866)(210:\lenssize)
        \centerarc[red,thick](-1.5,-3 * 0.866)(210:270:\lenssize)
        \connectionline[red,thick](-1.5,-3 * 0.866)(1.5,-3 * 0.866)(270:\lenssize)
        \centerarc[red,thick](1.5,-3 * 0.866)(270:330:\lenssize)
        \connectionline[red,thick](1.5,-3 * 0.866)(3,0)(330:\lenssize)
        \centerarc[red,thick](3,0)(330:390:\lenssize)
        \connectionline[red,thick](3,0)(1.5,3 * 0.866)(30:\lenssize)
        \node[red] at (0, 3.85) {\SI{99834}{\milli\metre\squared}};
    \end{scope}
\end{tikzpicture}
    \caption{Two possible arrangements for a hyperspectral camera array with 37 cameras. The orthogonal-spaced array needs more space.}
    \label{fig:eval_arrangements}
\end{figure}

\begin{table}[t]
	\footnotesize
	\centering
	\caption{Evaluation of a orthogonal-spaced hyperspectral camera in comparison to the proposed hexagonal shape, both with 37 cameras. The results are given by PSNR in dB / SSIM.}
	\label{tab:eval_rect}
        \begin{tabular}{l|cc}
                         & Orthogonal-spaced & Hexagonal   \\
        \hline
        Family house     & 29.32 / \textbf{0.962} & \textbf{29.42} / \textbf{0.962}  \\
        Medieval seaport & 43.80 / \textbf{0.988} & \textbf{43.95} / \textbf{0.988}  \\
        City             & 40.22 / \textbf{0.988} & \textbf{40.27} / \textbf{0.988}  \\
        Outdoor          & 28.59 / 0.915 & \textbf{28.82} / \textbf{0.917}  \\
        Forest           & 27.43 / 0.888 & \textbf{27.67} / \textbf{0.892}  \\
        Indoor           & 34.48 / 0.982 & \textbf{35.17} / \textbf{0.983}  \\
        Lab              & 27.86 / 0.966 & \textbf{28.51} / \textbf{0.969}  \\
        \hline
        Average          & 33.10 / 0.955 & \textbf{33.40} / \textbf{0.957}  \\
    \end{tabular}
\end{table}

In \fig\ref{fig:eval_arrangements}, two possible camera array arrangements are depicted, namely an orthogonal-spaced version and a hexagonal design.
In \tab\ref{tab:eval_rect}, these two camera array arrangements are evaluated.
For that, the synthetic database was also rendered using the orthogonal-spaced camera array.
Though the difference in terms of PSNR and SSIM between these two arrays is not huge, the hexagonal shape still outperforms the orthogonal-spaced design by 0.3 dB due to a smaller baseline between neighboring cameras.
Furthermore, in both cases the SSIM is better for the proposed hexagonal design.
This outperformance is consistent across all scenes.

Not only the performance of the hexagonal shape is better but also the space needed using the same baseline.
The area of the convex hull of the hexagonal design with a baseline of \SI{60}{\milli\metre} is \SI{99834}{\milli\metre\squared}, while the area of the convex hull of the orthogonal-spaced design with the same baseline calculates to \SI{116613}{\milli\metre\squared}.
Thus, the orthogonal-spaced design needs nearly 17\% more space.

\subsection{Limitations}

Despite the excellent registration quality, the hardware flexibility and the lower price point of such a hyperspectral camera array, this approach has some limitations regarding the scene that is captured.
While in far field every object has enough distance to the camera array, so that the disparity between different objects is small and therefore causes no problem, near-field scenes are different.
There, due to blocking objects, some cameras do not see all the objects, the center camera sees, since this camera has a different angle on the scene.
During reconstruction, the texture of the occluded object is inferred based on other similar objects visible in both cameras.
Therefore, this spectral information is only estimated and not recorded by any camera.
Moreover, as with every camera array, very close objects, that are only visible in some of the cameras, may cause problems, since the disparity cannot be properly estimated.
Furthermore, specular objects cause problems with the existing pipeline, since the reflections depend on the viewing angle.
Related to that, semi-transparent objects may induce problems because they need to be unmixed and have multiple disparities attached to it.
Note that these issues are implicit problems of any camera array.

\section{Real-world Data}
\label{sec:real_data}
\begin{figure*}
    \centering
    \vspace*{-1.5cm}
    \input{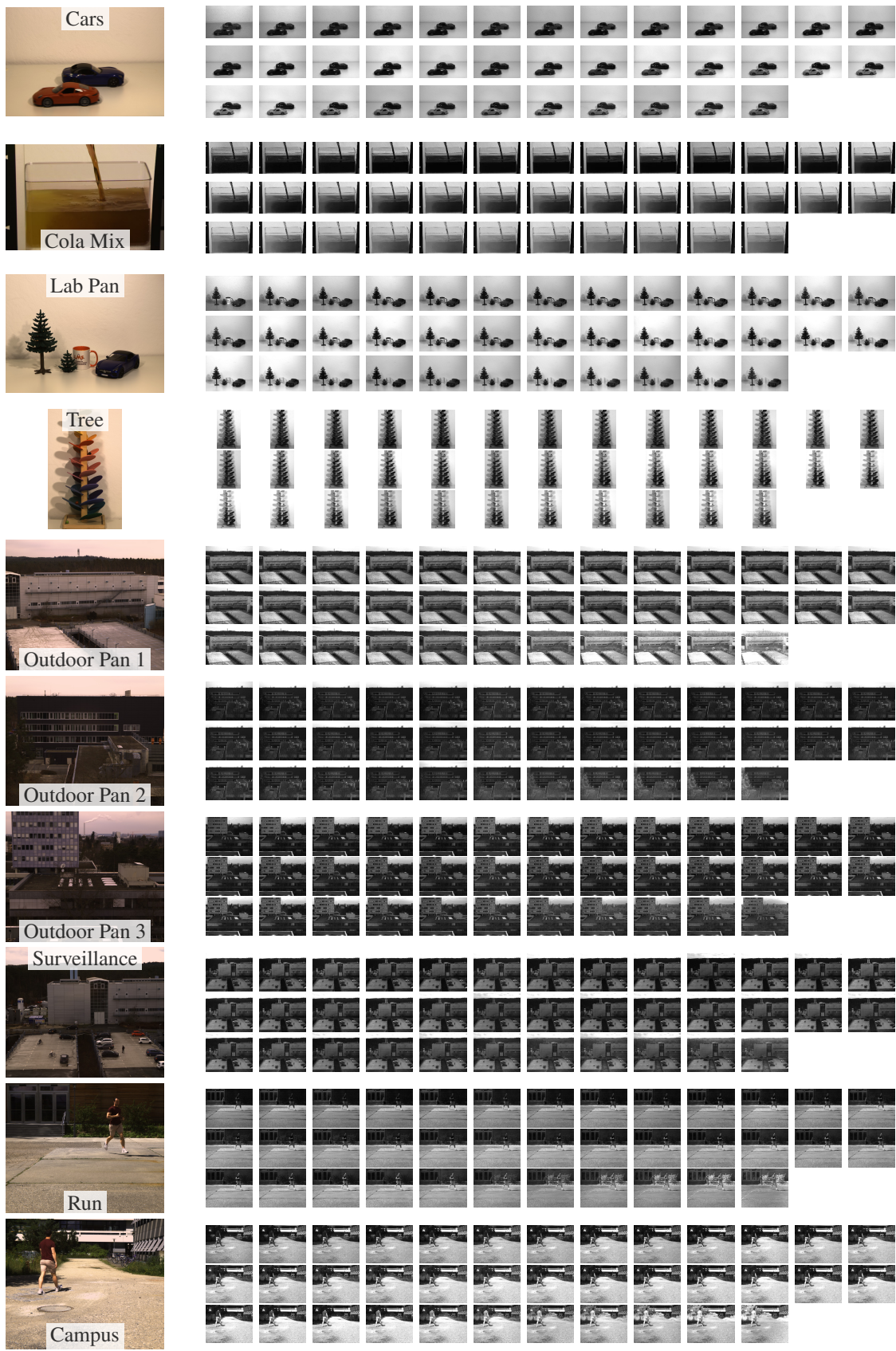}
    \caption{Real-world dataset. In total 10 scenes were recorded. On the left, you can see the corresponding RGB image rendered using standard CIE 1931 color curves. On the right next to it, all 37 corresponding channels are depicted from 400 nm to 760 nm in an ascending order.}
    \label{fig:database}
\end{figure*}

To provide high-resolution hyperspectral data for further research, the proposed array is used to capture high-resolution hyperspectral videos.
Moreover, the raw images as well as the calibration images are provided to test registration algorithms.

\subsection{Database}

The database consists of ten scenes, including four near-field lab scenes, four outdoor far-field scenes and two outdoor near-field scenes.
The basic facts are summarized in \tab\ref{tab:database}, while an RGB image and all corresponding 37 channels of one frame of each scene is shown in \fig\ref{fig:database}.
Note that the exposure time is the same for all cameras, since otherwise different spectral channels would contain different amounts of motion blur.
Hence, the gain is used to individually amplify the images to the full signal range.

This database contains a mix between higher spatial resolution and higher temporal resolution as well as different movements.
\textit{Cars} is a scene recorded in the lab, where two toy cars are moving in opposite direction on a table.
Though this scene is short with only 31 frames, the movement of the cars itself is fast paced.
Even with a low exposure time of only 5 ms, motion blur appears.
\textit{Cola Mix} is a scene where cola is poured into lemonade.
This scene is recorded with a high frame rate of 170 FPS while the resolution is kept at an acceptable level of 600 \texttimes\ 400.
\textit{Lap Pan} is a static scene in the lab, where the camera rotates to the right and then to the left again.
\textit{Tree} is a scene of a wooden toy tree with different colors at different layers.
Different colored balls are dropped onto this tree, which make their way down through the tree by dropping to the next layer.
Since the tree is taller than wide, the corresponding images have an aspect ratio much lower than 1.

\begin{table}[t]
	\footnotesize
	\centering
	\caption{Summary of all important facts (NF = near-field, FF = far-field) of the real-world high resolution hyperspectral database.}
	\label{tab:database}
	\begin{tabular}{@{\hspace*{0.1cm}}l|c@{\hspace*{0.25cm}}c@{\hspace*{0.25cm}}c@{\hspace*{0.25cm}}c@{\hspace*{0.25cm}}c@{\hspace*{0.25cm}}c@{\hspace*{0.1cm}}}
        Name            & Resolution            & Frame Rate    & Frames        & Exposure      & NF        & FF \\
        \hline
        Cars            & 1600 \texttimes\ 1100 & 30 FPS        & 31            & 5 ms          & \cmark    &           \\
        Cola Mix        & 600 \texttimes\ 400   & 170 FPS       & 3351          & 5 ms          & \cmark    &           \\
        Lab Pan         & 1600 \texttimes\ 1200 & 30 FPS        & 200           & 5 ms          & \cmark    &           \\
        Tree            & 800 \texttimes\ 1300  & 50 FPS        & 925           & 5 ms          & \cmark    &           \\
        Outdoor Pan 1   & 1124 \texttimes\ 924  & 23 FPS        & 233           & 10 ms         &           & \cmark    \\
        Outdoor Pan 2   & 1124 \texttimes\ 924  & 23 FPS        & 286           & 10 ms         &           & \cmark    \\
        Outdoor Pan 3   & 1124 \texttimes\ 924  & 23 FPS        & 316           & 10 ms         &           & \cmark    \\
        Surveillance    & 1124 \texttimes\ 924  & 23 FPS        & 1000          & 4 ms          &           & \cmark    \\
        Run             & 1132 \texttimes\ 928  & 22 FPS        & 380           & 1 ms          & \cmark    &           \\
        Campus          & 1132 \texttimes\ 928  & 22 FPS        & 900           & 1 ms          & \cmark    &           \\
	\end{tabular}
    \vspace*{-0.3cm}
\end{table}

\textit{Outdoor Pan 1}, \textit{Outdoor Pan 2} and \textit{Outdoor Pan 3} are outdoor scenes recorded from the 6th floor of a building.
The scenes themselves contain little movement, however, the camera rotates alternating to the right and left.
These scenes use two times two binning to capture more light for one pixel.
Similarly, \textit{Surveillance} was also shot from the 6th floor with two times two binning, however, the camera is static and the scene itself contains moving pedestrians, bicycles and cars.

Finally, \textit{Run} and \textit{Campus} are shot outside and contain near-field content.
Hence, the sun is illuminating both scenes.
Again, two times two binning is used and both scenes were recorded using 22 FPS and a very low exposure time of only 1 ms.
In \textit{Run}, a person first jogs from left to right, then runs from right to left, and finally jogs towards the camera.
Hence, this is a scene with complex and varying depths.
Finally, \textit{Campus} depicts a scene of a campus, where a person first walks away from the camera array.
In the end, he turns around and walks towards the camera.

\subsection{Applications}

The real-world database has numerous applications in different fields of hyperspectral signal processing.
First, it can be used for hyperspectral image and video coding.
The data rate this array produces can be calculated as
\begin{equation}
    37 \ \text{cameras} \cdot \frac{2448 \cdot 2048 \ \text{pixels}}{\text{camera and frame}} \cdot \frac{8 \ \text{bit}}{\text{pixel}} \cdot \frac{23 \ \text{frames}}{\text{s}} \approx 34.13 \ \frac{\text{Gbit}}{\text{s}}.
\end{equation}
Hence, a minute of hyperspectral video captured by HAHSI would produce 240 GB data, which is enormous to transmit.
Of course, classical video coders can be applied to pairs of three channels to simulate RGB videos, however, there is much more correlation to exploit.
Light spectra are typically smooth~\cite{sippel_spre_2020} and therefore nearby bands are highly correlated.
This fact can be exploited to reduce the bitrate massively.
Some multi- and hyperspectral image and video coders have been developed by~\cite{sippel_hyvid_2023, meyer_msc_2020}, but there is much more potential left.

Furthermore, this database can be used for spectral estimation~\cite{sippel_spre_2020}, where the goal is to estimation hyperspectral images from multispectral images or even RGB images.
Also, this database can serve as basis to extract principle spectra, e.g., for PCA-based hyperspectral algorithms~\cite{xudong_pca_2017}.
Moreover, a lot of work that has been done on RGB camera arrays can be extended to hyperspectral camera arrays, such as novel view synthesis~\cite{mildenhall_nvs_2019}.
Hyperspectral image and video denoising algorithms~\cite{xie_hyperspectral_2020} can use this database to evaluated their performance.
Finally, this data can also be used as evaluation basis for other approaches to hyperspectral imaging.
However, one has to keep in mind that the images of this database were also processed during the registration process.

\section{Conclusion}
\label{sec:conclusion}
In this paper, the hexagonal array for hyperspectral imaging, in short HAHSI, was introduced.
This array is able to record high-resolution hyperspectral videos using off-the-shelf industrial cameras, liquid lenses, and bandpass filters as main components.
Since all cameras are placed at different but well-defined positions, an image processing pipeline is suggested to register all images from each band.
To evaluate this approach against other popular approaches on high resolution hyperspectral imaging, a synthetic database was created, where a digital twin of the hexagonal array for hyperspectral imaging recorded seven scenes.
The evaluation revealed that HAHSI is able to outperform other popular approaches to hyperspectral video cameras.
In terms of PSNR, our method is able to achieve a gain of more than 2 dB on average over all seven scenes.
Furthermore, it was shown that an even better signal processing pipeline may easily add a gain of more than 1 dB by a better registration procedure.
Finally, a real-world high resolution hyperspectral video database was created, which can be beneficial in diverse applications and thus lead to better research in the field of hyperspectral imaging and hyperspectral signal processing.

\begin{backmatter}
 \bmsection{Funding} The authors gratefully acknowledge that this work has been supported by the Deutsche Forschungsgemeinschaft (DFG, German Research Foundation) under project number 491814627.
 \bmsection{Disclosures} The authors declare no conflicts of interest.
 \bmsection{Data availability} Data underlying the results presented in this paper are available in Dataset 1, Ref. \cite{hahsi-data}.
\end{backmatter}

%\bibliographyfullrefs{refs}
\bibliography{refs}

\end{document}